\documentclass[aps,pra,twocolumn,nofootinbib,showpacs]{revtex4-1}

\usepackage{graphicx}
\begin{document}

\title{Characterization of a cryogenic beam source for atoms and molecules}
\author{N. E. Bulleid}
\author{S. M. Skoff}
\author{R. J. Hendricks}
\author{B. E. Sauer}
\author{E. A. Hinds}
\author{M. R. Tarbutt}\email{m.tarbutt@imperial.ac.uk}

\affiliation{Centre for Cold Matter, Blackett Laboratory, Imperial College London, Prince Consort Road, London, SW7 2AZ, United Kingdom.}

\begin{abstract}
We present a combined experimental and theoretical study of beam formation from a cryogenic buffer gas cell. Atoms and molecules are loaded into the cell by laser ablation of a target, and are cooled and swept out of the cell by a flow of cold helium. We study the thermalization and flow dynamics inside the cell and measure how the speed, temperature, divergence and extraction efficiency of the beam are influenced by the helium flow. We use a finite element model to simulate the flow dynamics and use the predictions of this model to interpret our experimental results.
\end{abstract}

%\pacs{37.20.+j}

%37.20.+j 	Atomic and molecular beam sources and techniques

\maketitle

\section{Introduction}

A recent development in the formation of atomic and molecular beams is the use of cryogenically-cooled buffer gas sources \cite{Maxwell(1)05}. In these sources the molecules of interest are produced inside a cell containing a buffer gas of helium cooled to about 4\,K, and then they flow out through an aperture to form a cold, slow-moving beam. When the flow of helium through the cell is low, the source is effusive and the speed of the beam scales as $\sqrt{T/M}$, where $T$ is the temperature and $M$ is the mass of the molecule. For heavy molecules produced at low temperature, the speed is particularly low. These effusive beams have a relatively low intensity, but this can be greatly increased by increasing the helium flow rate so that the molecules are caught up in the flow and efficiently swept out of the cell. This also increases the speed of the beam, which eventually reaches the supersonic speed of helium, scaling as $\sqrt{T/m_{\text{He}}}$ where $m_{\text{He}}$ is the mass of a helium atom. For low temperatures, this speed is still lower than other supersonic sources. In most experiments the sources have been operated in the intermediate flow regime, neither effusive nor fully supersonic, where the beam is intense, cold, slow and collimated. These molecular beam sources are increasingly being used for precise spectroscopy and tests of fundamental physics \cite{Vutha(1)10}, and for experiments on molecule deceleration \cite{Bulleid(1)12} and laser cooling \cite{Shuman(1)10, Barry(1)12, Hummon(1)12}.

There have been several previous studies of beam formation from cryogenic buffer gas sources\cite{Maxwell(1)05, Patterson(1)07, Patterson(1)09, Barry(1)11, Hutzler(1)11, Lu(1)11}, and these are reviewed in \cite{Hutzler(1)12}. Here we present results from a buffer gas source where atoms and molecules are produced by laser ablation of a target inside the cell. We study the dynamics inside the cell and characterize the properties of the beam. We use a model of the flow dynamics to help understand our observations.

\section{Apparatus}

\begin{figure}[tb]
	\begin{center}
		\includegraphics[width=0.75\linewidth]{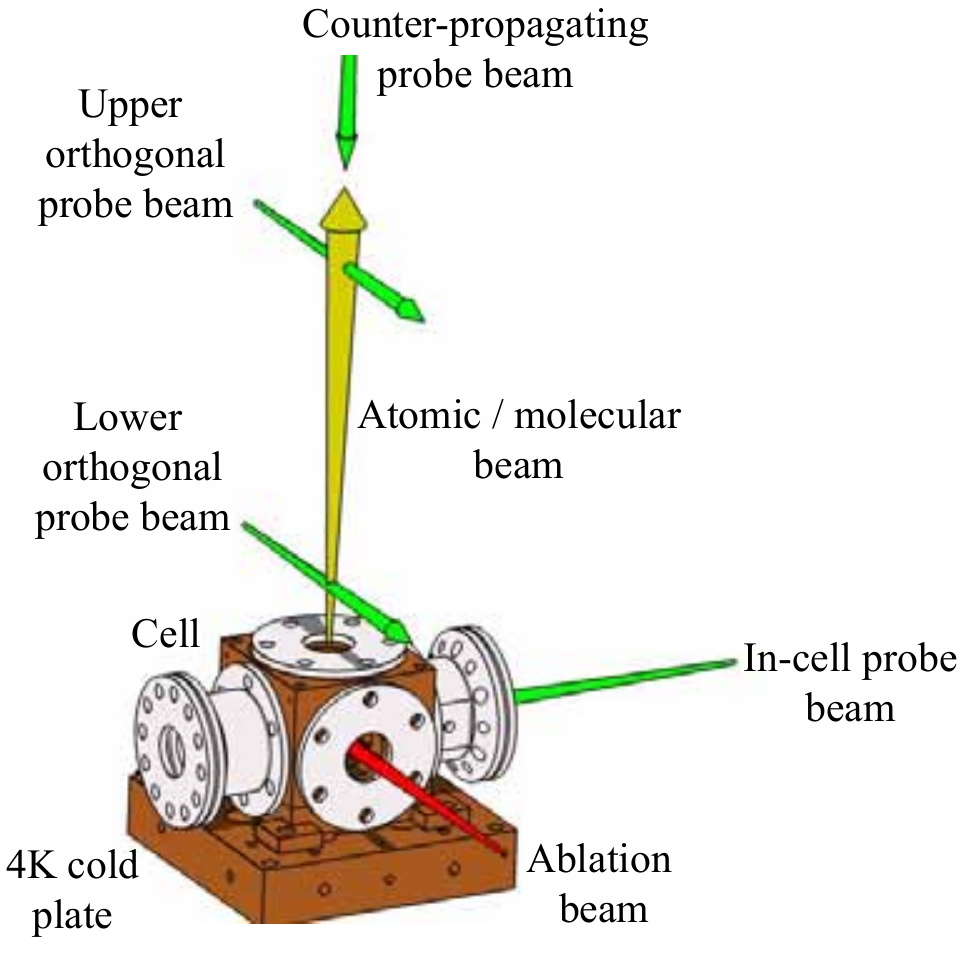}
	\end{center}
	\caption{Setup of the apparatus}
	\label{fig:Setup}
\end{figure}

Figure \ref{fig:Setup} shows our experimental set-up. The copper cell is a cube with an internal edge of 30\,mm, and it is attached to the cold stage of a two-stage cryocooler, where the temperature is measured to be 4\,K. On three sides of the cell there are 16\,mm diameter windows for optical access. Helium gas, pre-cooled to 4\,K, flows into the cell through a 2\,mm hole, from a tube whose internal diameter is 2.5\,mm. The flow of helium is controlled by two flow meters in parallel, which have ranges of 0-20\,sccm (standard cubic centimeters per minute) and 0-100\,sccm respectively. The beam flows out of the cell through an aperture on the top. We have experimented with two different slit-shaped apertures, 0.75\,mm x 4\,mm  and 1\,mm x 8\,mm. In the direction of the beam, both are 1\,mm deep. Outside the cell the helium is pumped away by charcoal sorption pumps. Aluminium radiation shields are anchored to the first stage of the cryo-cooler whose temperature is 40\,K. The apparatus is housed in a vacuum chamber evacuated by a turbomolecular pump. A vacuum gauge far from the cold region reads a pressure of $10^{-8}$ mbar when there is no gas flow, rising to about $10^{-6}$\,mbar under typical running conditions.

Yb is introduced into the cell by laser ablation of a Yb metal target placed inside the cell opposite one of the windows. The ablation pulses have a wavelength of 1064\,nm, a duration of 8\,ns, a spot size at the target of 1.5\,mm, and pulse energies between 50 and 150\,mJ. The Yb is detected by absorption or by laser-induced fluorescence spectroscopy on the $\text{6s}^{2}\,^{1}S_{0} \rightarrow \text{6s6p} ^{3}P_{1}$ transition of Yb at 556\,nm using a continuous-wave frequency-doubled fibre-laser. For absorption spectroscopy of the atoms inside the cell, the probe beam is expanded to uniformly fill the 16\,mm diameter windows of the cell. After exiting from the vacuum chamber, this probe beam is split into two parts, one detected by a CCD-camera, and the other focussed onto a photodiode. The atomic beam exiting from the cell is detected by laser-induced fluorescence (LIF) 23\,mm and 132\,mm from the exit aperture. At each detector the atoms can be excited by one of two laser beams, one orthogonal to the atomic beam, and the other counter-propagating to the atomic beam. The resulting fluorescence is collected by a spherical mirror and an aspheric lens and imaged onto a photomultiplier tube (PMT). The counter-propagating probe laser is focussed through the exit aperture of the cell to minimize background laser scatter at the lower PMT. At the lower detector, the orthogonal laser beam is detected by a photodiode to obtain an absorption signal at this point.

Data are taken at a repetition rate of 5\,Hz. For each shot we record the signals from the two photodiodes and the two PMTs, each at a sample rate of 100\,kHz, and we record a single picture from the CCD camera which is gated on for a specified time window of duration 34 $\mu$s. The Q-switch of the ablation laser marks the zero of time. To obtain a spectrum the frequency of the probe laser is stepped between shots and measured by a wavemeter with an accuracy of 600\,MHz, and by a low finesse Fabry-Perot cavity with a free spectral range of 150\,MHz.

\section{Model}

We simulate the flow of the buffer gas and the injected atoms and molecules using finite element modelling software\footnote{Star-CCM+ from CD-adapco} to solve the Navier-Stokes equations. This approach is valid in the regime of continuum flow which, in our case, holds inside the cell and for a short distance outside the cell. In the model a volume, corresponding to the internal volume of our cell, contains a gas that obeys the Peng-Robinson equation of state \cite{Peng(1)76} and has the physical properties of helium. The walls are held at 4\,K. The pressure far from the outlet is set to zero and the mass flow at the inlet is a parameter fixed for a given simulation in the range from 1--60\,sccm. The flow is assumed to be non-turbulent. The cell is initally filled with gas at a uniform pressure equal to the expected pressure for the mass flow used. The model is then
iterated to a steady state which reveals the helium flow through the cell. This is then used to model the dynamics of the Yb in the cell, which is governed by the transport equation,
\begin{equation}
	\frac{\partial n_{\text{at}}}{\partial t} + \nabla \cdot (\vec{v} n_{\text{at}}) = \nabla \cdot (D \nabla n_{\text{at}}),
\end{equation}
where $n_{\text{at}}$ is the scalar field representing the number density of Yb, $\vec{v}$ is the velocity of the helium
and $D$ is the diffusion coefficient. The
interaction between the helium and the Yb is contained in the diffusive term depending on the diffusion coefficient, and therefore the Yb-He collision properties, and in the advective term which depends only on the velocity of the helium. The source term
is omitted as we are interested in the evolution of an initial
distribution. The transport equation was solved for 82\,ms using a time step of 0.1\,ms. In the experiments, the helium density is high enough that the ballistic expansion of the Yb away from the ablation point is quickly arrested, typically quite close to the target. So we use a localized distribution of Yb as the initial condition, and vary the position and size of this distribution to see the effect of these initial parameters.

\section{Results}
\subsection{Flow through the cell.\label{Sec:Flow}}

\begin{figure}[tb]
	\begin{center}
		\includegraphics[width=\linewidth]{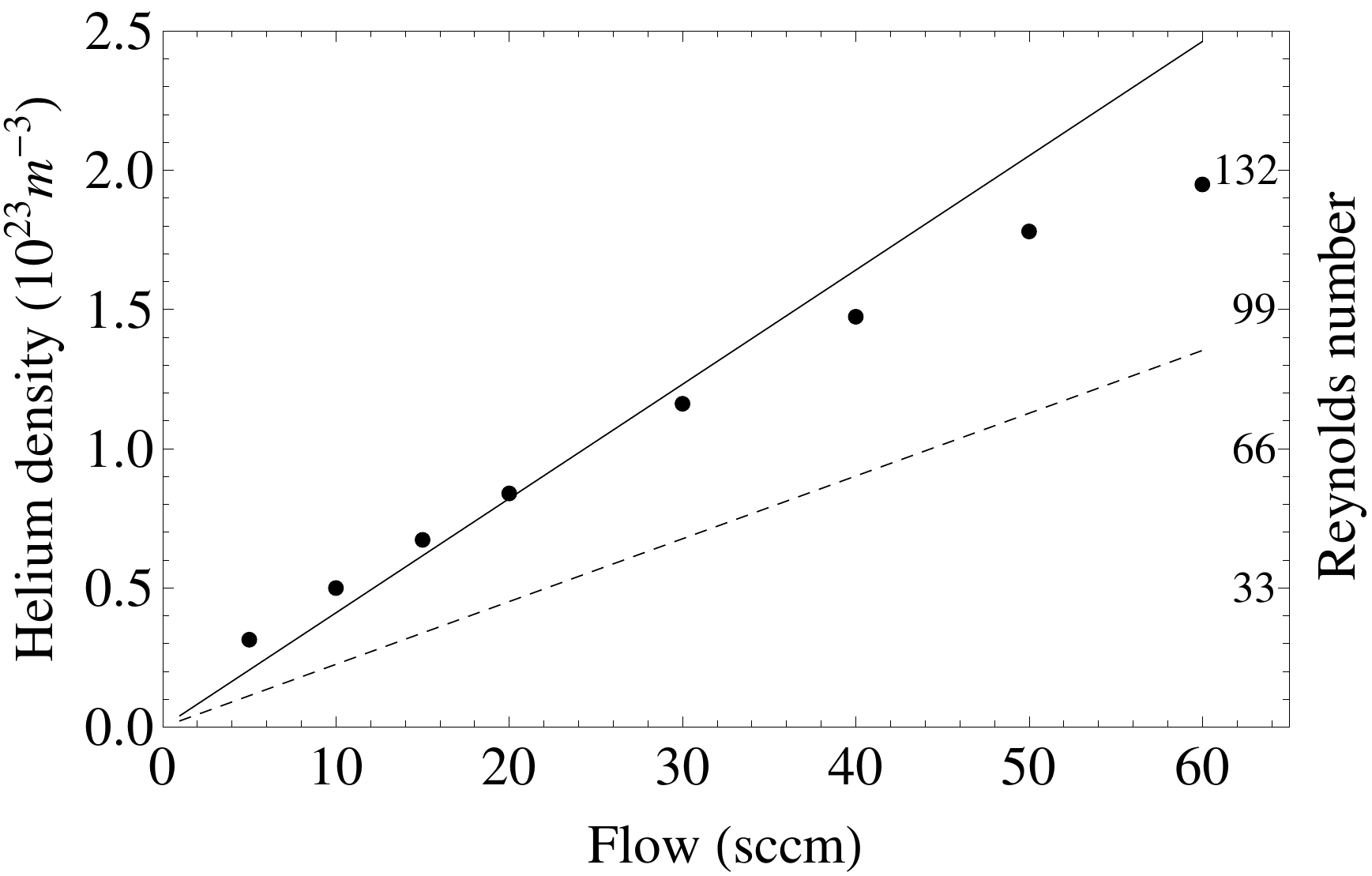}
	\end{center}
	\caption{Helium number density in the cell as a function of the flow. Solid and dashed lines show the effusive and hydrodynamic limits, and points show the values obtained numerically. The right hand axis gives the approximate Reynolds number for a 0.75\,mm wide slit.}
	\label{fig:DensityVsFlow}
\end{figure}

In the effusive regime, the flow rate, in particles per second, out of a thin aperture of area $a$, is
\begin{equation}
\dot{N}=\frac{n \bar{v} a}{4}.
\label{equ:flowrateeff}
\end{equation}
The maximum hydrodynamic flow rate is higher and is given by \cite{Pauly(1)00}
\begin{equation}
\dot{N}=n \bar{v} a\frac{\sqrt{\pi}}{2} \Big(\frac{\gamma}{\gamma+1}\Big)^{1/2}\Big(\frac{2}{\gamma+1}\Big)^{1/(\gamma-1)}.
\label{equ:flowratehyd}
\end{equation}
Here, $\bar{v}$ is the average thermal velocity of the buffer gas atoms, $n$ is the number density, and $\gamma$ is the specific heat ratio, $\gamma = 5/3$ for a monoatomic gas such as helium. When $\gamma = 5/3$ this maximum flow rate is 1.8 times the effusive flow rate given by equation (\ref{equ:flowrateeff}). Figure \ref{fig:DensityVsFlow} shows the relationship between the number density and the flow for helium at 4\,K in the effusive and the fully hydrodynamic limits. The numerical simulation gives more exact values for the number density in the cell at various flow rates, shown by the points in figure \ref{fig:DensityVsFlow}. As expected, the values are close to the effusive limit at low flow and converge towards the hydrodynamic limit at high flow. It is common to characterize flow regimes using the Reynolds number, which for the helium escaping through the aperture is proportional to the ratio of the smallest aperture dimension to the helium mean free path. This is proportional to the helium number density in the cell. The right hand axis in figure \ref{fig:DensityVsFlow} converts the left hand axis to an approximate value for the Reynolds number in the case of the 0.75\,mm wide slit, where we have taken the helium collision cross-section to be $2.6 \times 10^{-19}$\,m$^{2}$ \cite{Hasted(1)72}. For all the experiments reported here, the flow was in the range 1-60 sccm, and the Reynolds number in the range 1-100. This falls in the regime intermediate between effusive and fully hydrodynamic. There are some collisions in the region of the aperture which change the properties of the beam from those present in the cell, but not so many that the beam is fully hydrodynamic.

\begin{figure}
	\begin{center}
		\includegraphics[width=\linewidth]{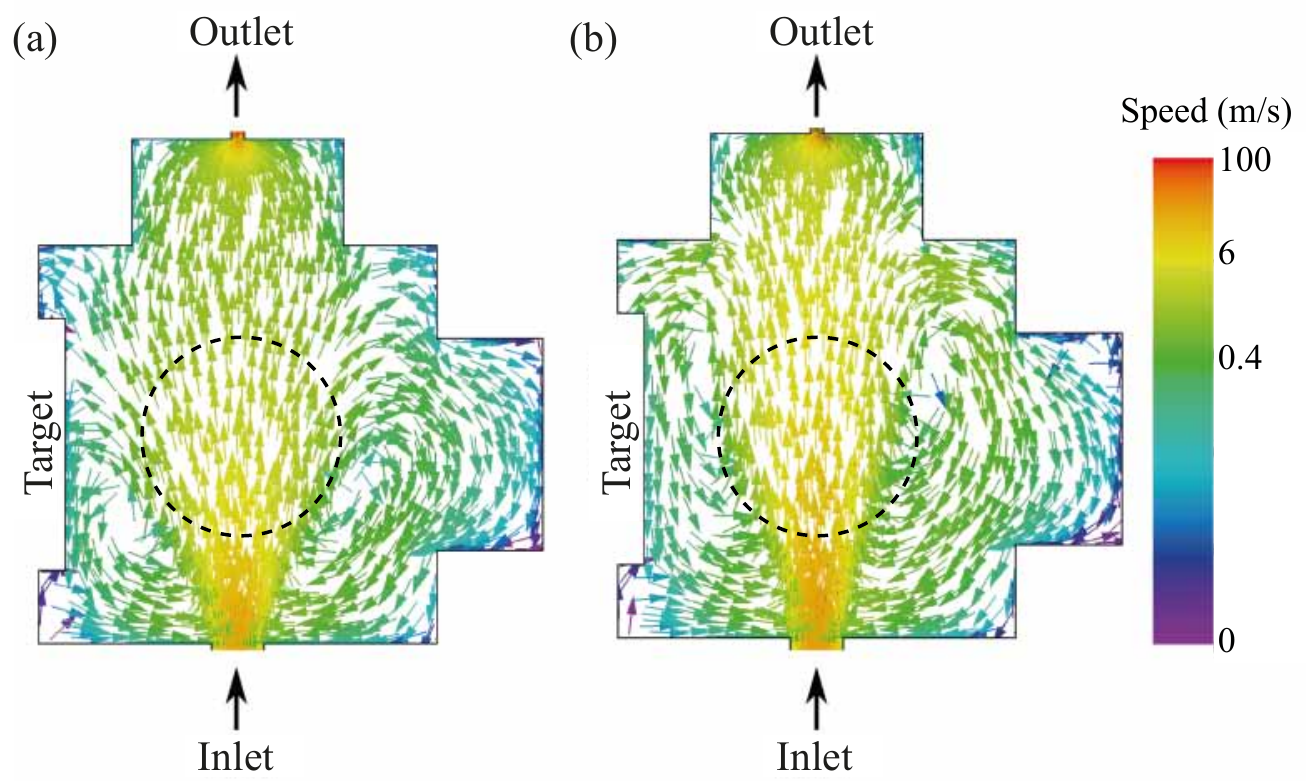}
	\end{center}
	\caption{Cross-section through the centre of the cell showing the simulated flow of helium gas. (a) 20\,sccm and (b) 40\,sccm. The dashed circle shows the position of the observation window that we use to probe the atoms and molecules within the cell.}
	\label{fig:gasFlow}
\end{figure}

Figure \ref{fig:gasFlow} shows the simulated flow of helium through the cell for flow rates of 20 and 40\,sccm. We have taken a slice through the centre of the cell that contains the inlet tube and the outlet aperture. The windows and the ablation target complicate the internal shape of the cell. The simulation shows a column of helium flowing from inlet to outlet and the formation of vortices on both sides of the central column. As the flow rate is increased the speed of the helium through the cell increases, the central column narrows and the vortices grow in size. For high flows, the helium density is high enough that the atoms and molecules formed by laser ablation are confined relatively close to the target. From here, they have to diffuse into the column of flow if they are to be swept out of the cell efficiently. We expect a large fraction to be trapped in the vortices, or to be formed in regions where they diffuse to the walls because the helium flow is too slow. This will limit the fraction that can be extracted out of the cell.

It is informative to estimate the characteristic time scales for various transport processes within the cell. Let us do that for a typical flow rate of 40\,sccm. On the centreline of the flow column, where the flow is fastest, the time taken for the helium to flow from the inlet to the outlet is found from the simulation to be 6\,ms. Atoms produced inside the flow column will be swept out of the cell on this time scale. The characteristic timescale for all the helium to leave the cell is
\begin{equation}
\tau_{\text{p}}=\frac{n V}{\dot{N}},
\label{equ:pump}
\end{equation}
where $V$ is the cell volume and $n$ is the mean number density of helium. For our parameters, $\tau_{\text{p}} = 220$\,ms.
Turning to diffusion, the lowest order diffusion mode in a cubic cell has a time constant of \cite{Hasted(1)72}
\begin{equation}
\tau_{\text{D}} = \frac{L^2}{3\pi^2 D},
\label{equ:diff}
\end{equation}
where $D$ is the diffusion coefficient which, for diffusion of one gas into another, can be expressed as \cite{Chapman(1)16}
\begin{equation}
D=\frac{3}{16} \left(\frac{2 \pi k T}{\mu}\right)^{1/2}\frac{1}{n \bar{\sigma}_{\text{D}}}.
\label{equ:diffcoefficient}
\end{equation}
Here, $\bar{\sigma}_{\text{D}}$ is the thermally averaged diffusion cross-section, $T$ is the temperature, and $\mu$ is the reduced mass of the two species. For our cell at 4\,K, taking $\bar{\sigma}_{\text{D}}=10^{-18}$\,m$^{2}$ as typical, this time constant is $\tau_{\text{D}} = 78$\,ms.

The laser ablation process produces a non-uniform initial distribution of atoms inside the cell. The subsequent dynamics depends strongly on their initial position. If they are inside the central flow column they will be swept out of the cell on a time scale of about 6\,ms. They are not swept out of the cell quickly from any other region since $\tau_{\text{p}} > \tau_{\text{D}}$. Instead they either diffuse to the walls where they are lost, or they diffuse slowly into the flow column and are then swept out.

\begin{figure}
	\begin{center}
		\includegraphics[width=\linewidth]{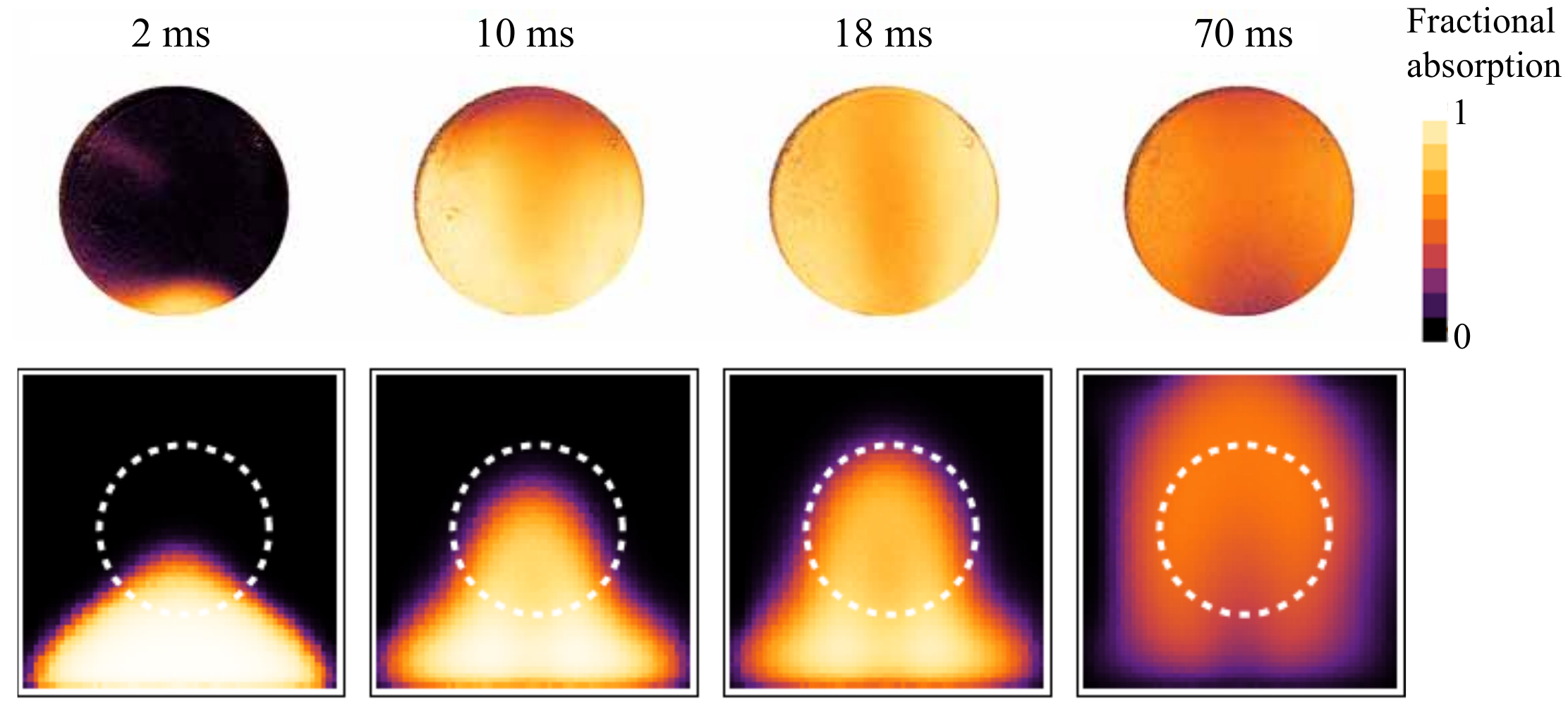}
	\end{center}
	\caption{Top row: Absorption images taken at various times, showing the evolution of Yb atoms through the cell. The flow rate is 20\,sccm and the ablation energy is 24\,mJ. The ablation target is on the left hand side and the outlet is at the top. Bottom row: Simulated absorption images at these same times and for the same flow rate. The initial density distribution is the top section of a sphere of radius 15\,mm centred 5\,mm below the bottom of the cell, and the initial density is chosen so that the degree of absorption approximately matches that observed in the experiment. The circular region in the simulated images corresponds to the viewing region in the experiment.}
	\label{fig:absorptionImages}
\end{figure}

In the experiment we can watch the evolution of the Yb density inside the cell by taking absorption images at various times after firing the ablation laser. Figure \ref{fig:absorptionImages} shows a series of such absorption images. The circular region is 16\,mm in diameter, the diameter of the viewports, which are completely filled by the probe laser beam. Each image corresponds to one shot of the experiment, under nominally identical conditions. The ablation plume deposits the atoms into a localized region of the cell which depends on the ablation conditions. In the example shown here, most of the atoms are initially near the bottom of the cell, as we see in the image taken at 2\,ms. After 10\,ms the atoms have filled most of the viewing area, and a column of lower density is visible in the centre, corresponding to the central flow column shown in figure \ref{fig:gasFlow}. The Yb density is lower here because the atoms are rapidly swept out of this region by the helium flow. This is even more pronounced in the image at 18\,ms. On either side of the flow column the atoms are caught in the vortices and remain in these regions for long periods of time, as we see in the image taken at 70\,ms.  Figure \ref{fig:absorptionImages} also shows simulated absorption images at the same times as the experimental images. These are produced by finding the density on a three dimensional grid of points and then summing over the grid points in the direction of the probe laser beam. Starting with atoms near the bottom of the cell, we see that the simulation predicts all the main features observed in the experiment. Examining absorption images at higher flow rates we see that the region of lower density narrows as the flow increases because the flow channel narrows, as shown in figure \ref{fig:gasFlow}. We also see that the regions of high Yb density are more pronounced at higher flows because the atoms are more strongly confined to the vortices.

\begin{figure}
	\begin{center}
		\includegraphics[width=0.8\linewidth]{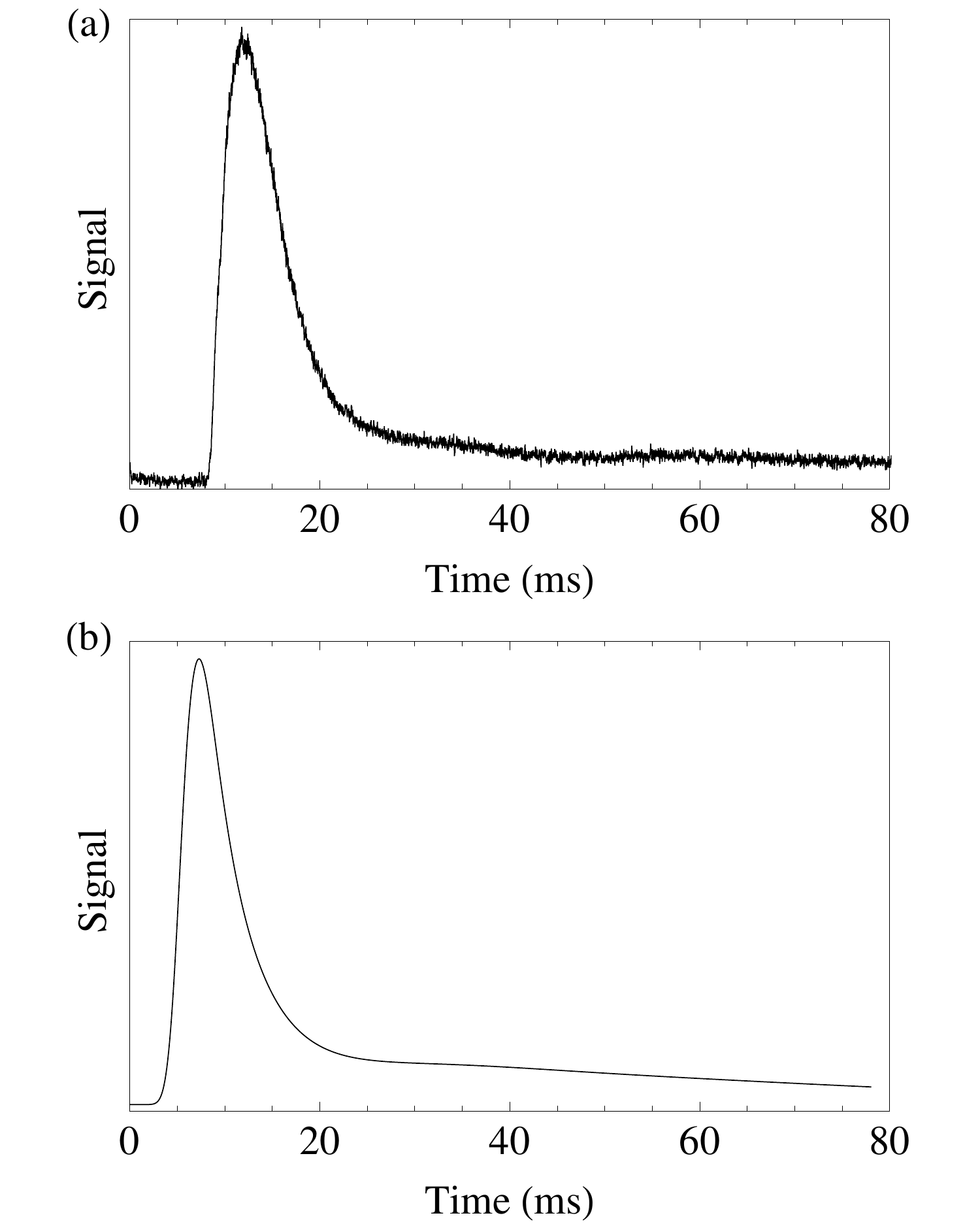}
	\end{center}
	\caption{Yb signal at the lower detector as a function of time since ablation, for a flow of 40\,sccm. (a) Measured, using an ablation energy of 24\,mJ. (b) Simulated.}
	\label{fig:timeProfiles}
\end{figure}

Figure \ref{fig:timeProfiles} shows how the measured and simulated Yb signal 23\,mm downstream of the exit aperture changes with time. For this case, the flow rate is 40\,sccm. The measured profile consists of a peak centred at 12\,ms with a width of approximately 6\,ms, and a long, low-amplitude tail of atoms arriving at later times. The simulated time-of-arrival profile is similar to the measured one and helps us to understand the shape of the profile. The peak is due to the atoms that are initially inside the central flow column. They are swept out of the cell on a time scale that is close to the expected 6\,ms. The long tail in the profile is due to molecules that are produced in other regions of the cell and are only swept out once they have diffused into the flow column. The decay time of this long tail is the characteristic diffusion time in the cell, $\tau_{D}$. The entire simulated profile is shifted in time relative to the measured one and we find that this shift depends on the initial distribution of ablated atoms. If the initial distribution is further from the exit aperture the whole profile is shifted to later times. We also see in the experiment that the peak arrival time depends on the position of the ablation beam on the target, and on the ablation laser power. With higher ablation powers molecules start arriving at the detector at earlier times.

\begin{figure}
	\begin{center}
		\includegraphics[width=0.9\linewidth]{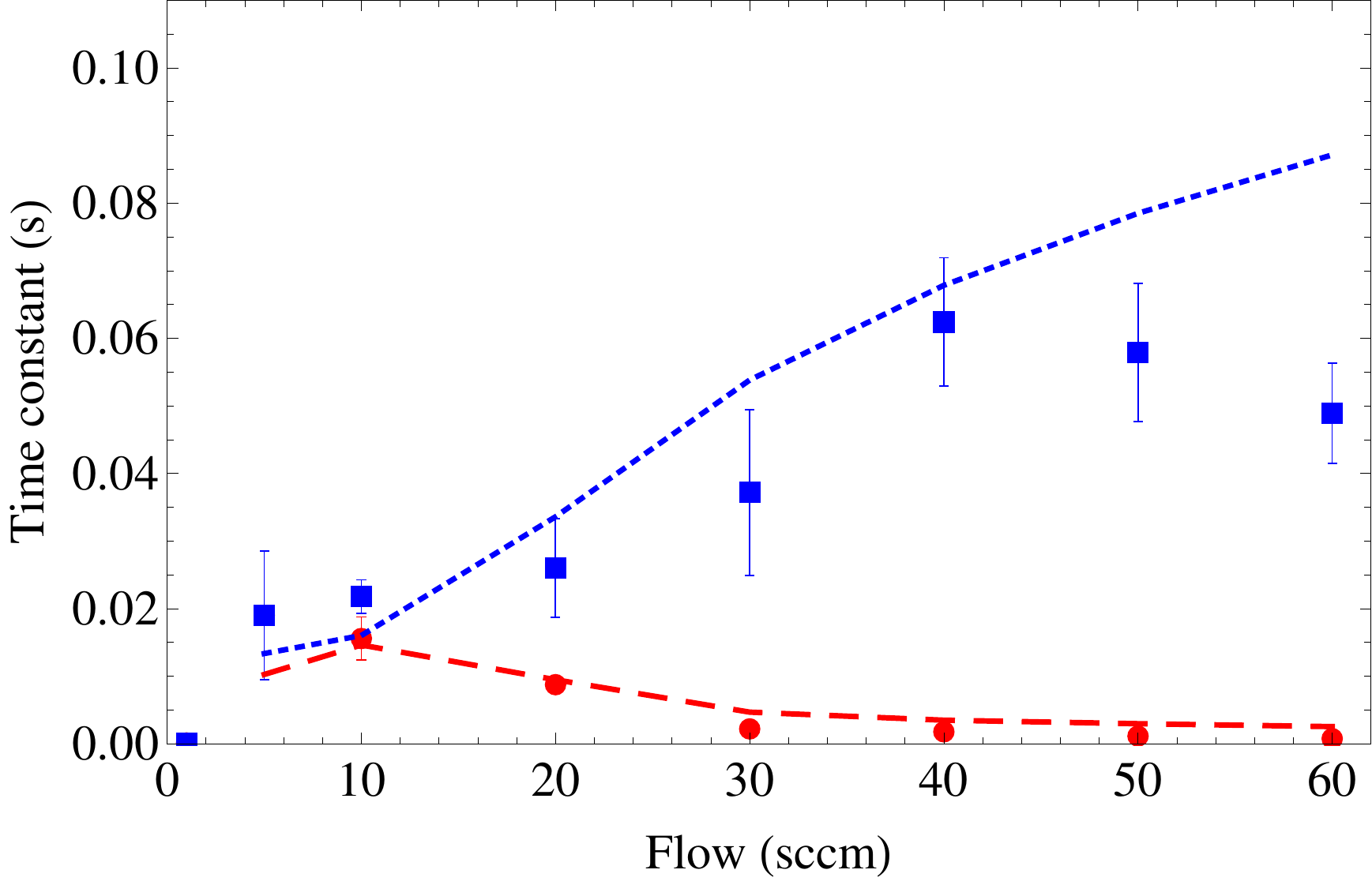}
	\end{center}
	\caption{Time constants of a double exponential decay model fit to the decaying part of the time profiles obtained at various flows (e.g. Fig.\,\ref{fig:timeProfiles}). Circles: experiment, short peak. Squares: experiment, long tail. Dashed line: simulation, short peak. Dotted line: simulation, long tail.}
	\label{fig:timec}
\end{figure}

The shape of the time profile depends on the helium flow rate through the cell. To quantify this, we fit a double exponential decay model to the decaying part of the arrival-time profiles, for both the measured and simulated data. Figure \ref{fig:timec} shows the two time constants obtained from these fits as a function of the flow rate. As the flow is increased the initial peak becomes shorter because the molecules that are in the flow column are swept out more rapidly. By contrast, the duration of the long tail increases with increasing flow because the diffusion time increases as the helium density increases. The model and experiment show the same trends. We also see that, as the flow rate increases, the amplitude of the short peak decreases relative to the amplitude of the long tail. This is because the flow column becomes narrower with increasing flow. At high flows the long time constant stops increasing and deviates from the prediction of the simulation. At these high flows the observed arrival time distribution has a more complicated shape than the simple double exponential decay model, and so our model is no longer a good fit to the data. This is most evident at 60\,sccm where we observe a small, broad bump in the arrival time distribution centred near 40\,ms, probably because the initial distribution of atoms is confined to a small region and several diffusion modes play a role.

\subsection{Temperature in the cell.}\label{Sec:TemperatureInside}

\begin{figure}[tb]
	\begin{center}
		\includegraphics[width=0.9\linewidth]{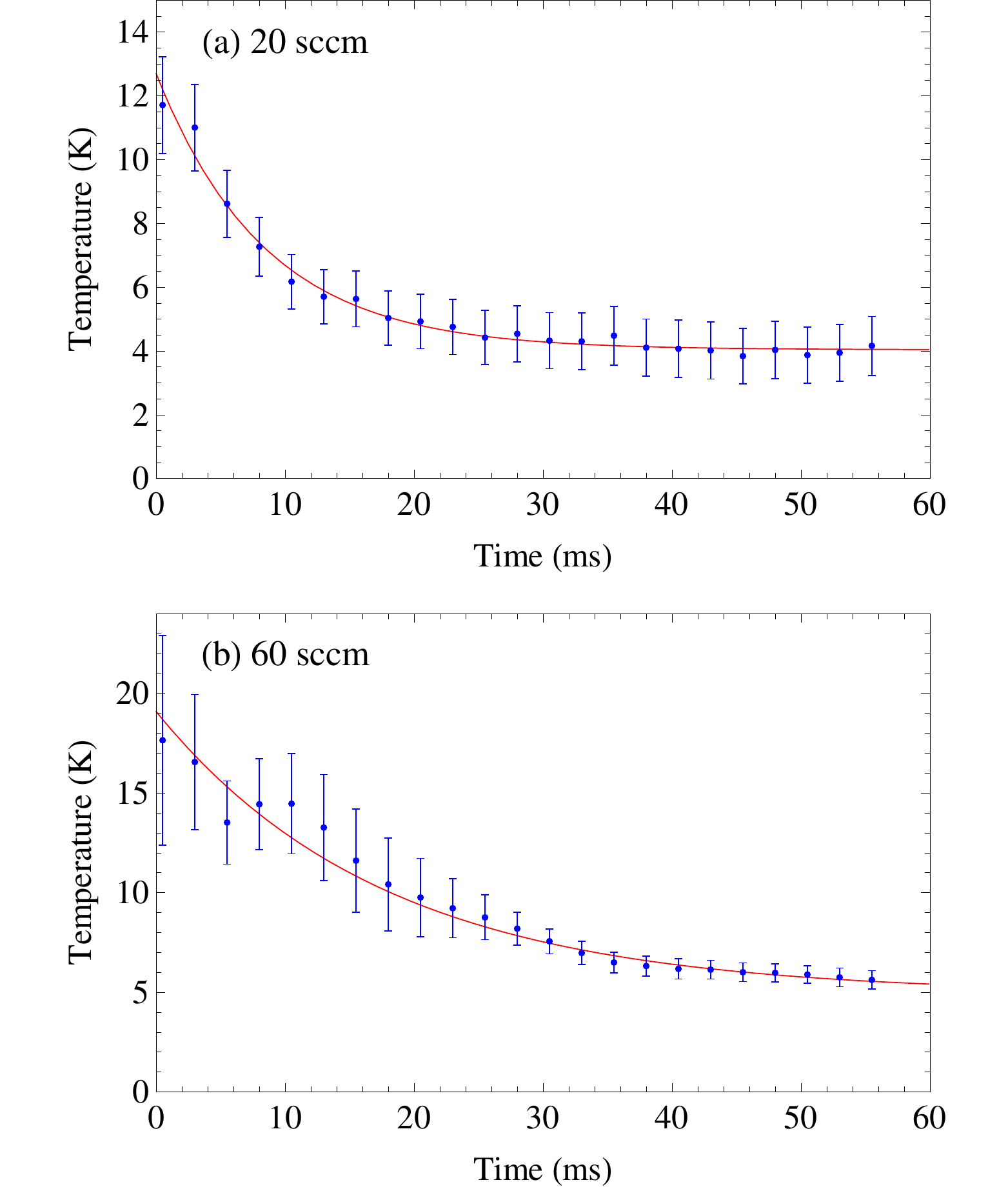}
	\end{center}
	\caption{Temperature of atoms inside the cell as a function of time after the ablation pulse for flows of (a) 20\,sccm and (b) 60\,sccm. The ablation energy is 94\,mJ. Lines are fits to an exponential decay model. The error bars shown are the ones found from fits to Voigt profiles and appear non-statistical because the data points are correlated.}
	\label{fig:temperature}
\end{figure}

The Yb inside the cell is detected by absorption spectroscopy. The optical density is obtained from the logarithm of the fractional absorption and the spectral line is then fit to a Voigt function. The fits show that the dominant line broadening mechanism is Doppler broadening, though with a significant contribution from pressure broadening at higher flows. From the width of the Gaussian component we find the temperature of the atoms. Figure \ref{fig:temperature} shows the temperature as a function of time since the ablation pulse, for an ablation energy of 90\,mJ and for two different flows, 20 and 60\,sccm. The initial temperature is raised above the cell temperature and then decreases on a timescale of order 10\,ms. Because this is much longer than the Yb-He thermalization time, the Yb temperature is measuring the helium temperature. Ablation of the target heats the helium and the time constant observed is the time taken for the deposited heat to diffuse to the cell walls \cite{Skoff(1)11}. The lines in figure \ref{fig:temperature} are fits to a single exponential decay model, $T(t)=T_{f} + (T_{i} - T_{f})\exp(-t/\tau)$, where $T_{i}$ and $T_{f}$ are the initial and final temperatures, and $\tau$ is the thermalization time constant. This model fits well for all values of the flow. From these fits the final temperature is found to be $4.7 \pm 1$\,K irrespective of the flow, where the uncertainty here is the standard deviation of the results at different flows. The initial temperature is $14 \pm 4$\,K with a slight trend towards higher temperatures at higher flow.

\begin{figure}[tb]
	\begin{center}
		\includegraphics[width=0.8\linewidth]{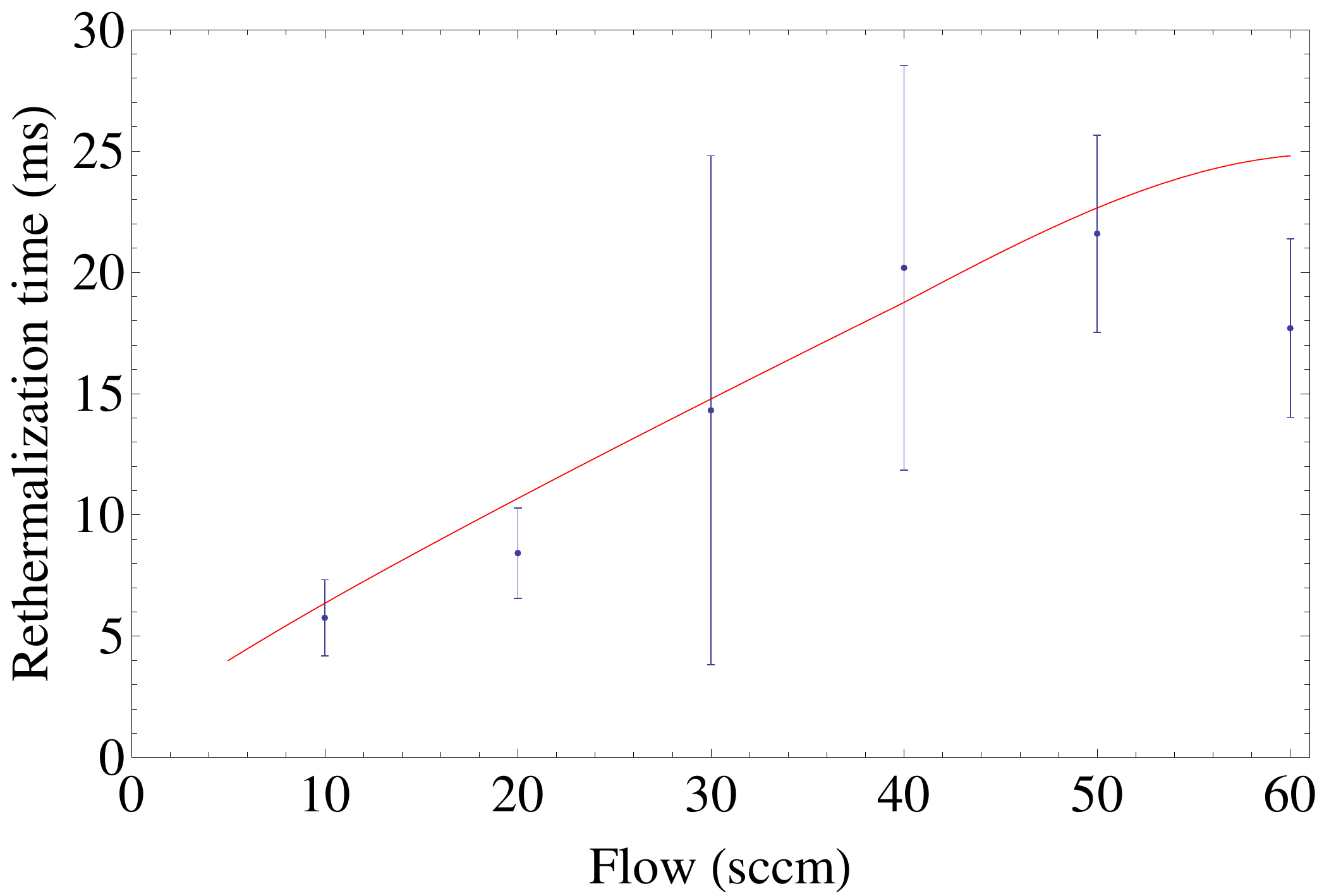}
	\end{center}
	\caption{Time constant for rethermalization following laser ablation with an energy of 94\,mJ. The line shows the time constant for the lowest-order heat diffusion mode in a 3\,cm cubic cell of helium.}
	\label{fig:rethermalization}
\end{figure}

Figure \ref{fig:rethermalization} shows the rethermalization time constant, $\tau$, as a function of flow. The time constant is higher for higher flows because the heat takes longer to diffuse to the walls when the helium density is higher. In a cubic cell of side length $a$, the thermal diffusion time constant for the lowest-order diffusion mode is $\tau_D = a^{2}/(3\pi^{2}\alpha)$ where $\alpha$ is the thermal diffusivity and is given by $\alpha=k/C_p n$. Here, $k$ is the thermal conductivity ($k=8.25\times10^{-3}$\,W\,m$^{-1}$\,K$^{-1}$ at 4\,K \cite{Vargaftik(1)77}), $C_p$ is the heat capacity at constant pressure, and $n$ is the number density. The calculated diffusion time as a function of flow is shown by the line in figure \ref{fig:rethermalization}, where we have used the relationship between number density and flow shown by the points in figure \ref{fig:DensityVsFlow}. We find that this simple model of heat diffusion to the cell walls fits well to our measurements.

The widths of the Lorentzian components of the Voigt fits are in the range 6--22\,MHz. These widths are much greater than the natural width of the transition and so we assume they are entirely due to pressure broadening. The widths increase with flow as expected. From the full width at half maximum $\gamma$ (in angular frequency units), we obtain the pressure broadening cross-section for the 556\,nm Yb transition in a helium buffer gas, $\sigma_b = \gamma/(n_{\text{He}} \bar{v})$, where $n_{\text{He}}$ is the helium number density and $\bar{v}=158$\,m/s is the mean speed of the helium atoms at 4.7\,K. From the fits we find a consistent result for $\sigma_b$ which is independent of the flow: $\sigma_b = (1170 \pm 50)\times 10^{-20}$\,m$^{2}$. Pressure broadening of this Yb transition in the presence of a helium buffer gas at 700\,K was investigated in reference \cite{Kimball(1)99}. Their results give a pressure broadening cross-section of $\sigma_b = (197 \pm 5)\times 10^{-20}$\,m$^{2}$ at 700\,K. The increasing cross-section with decreasing temperature is consistent with other observations of pressure broadening at low temperatures, e.g. \cite{Willey(1)88}

\subsection{Speed of the beam.}
\label{Sec:Velocity}

\begin{figure}[tb]
	\begin{center}
		\includegraphics[width=\linewidth]{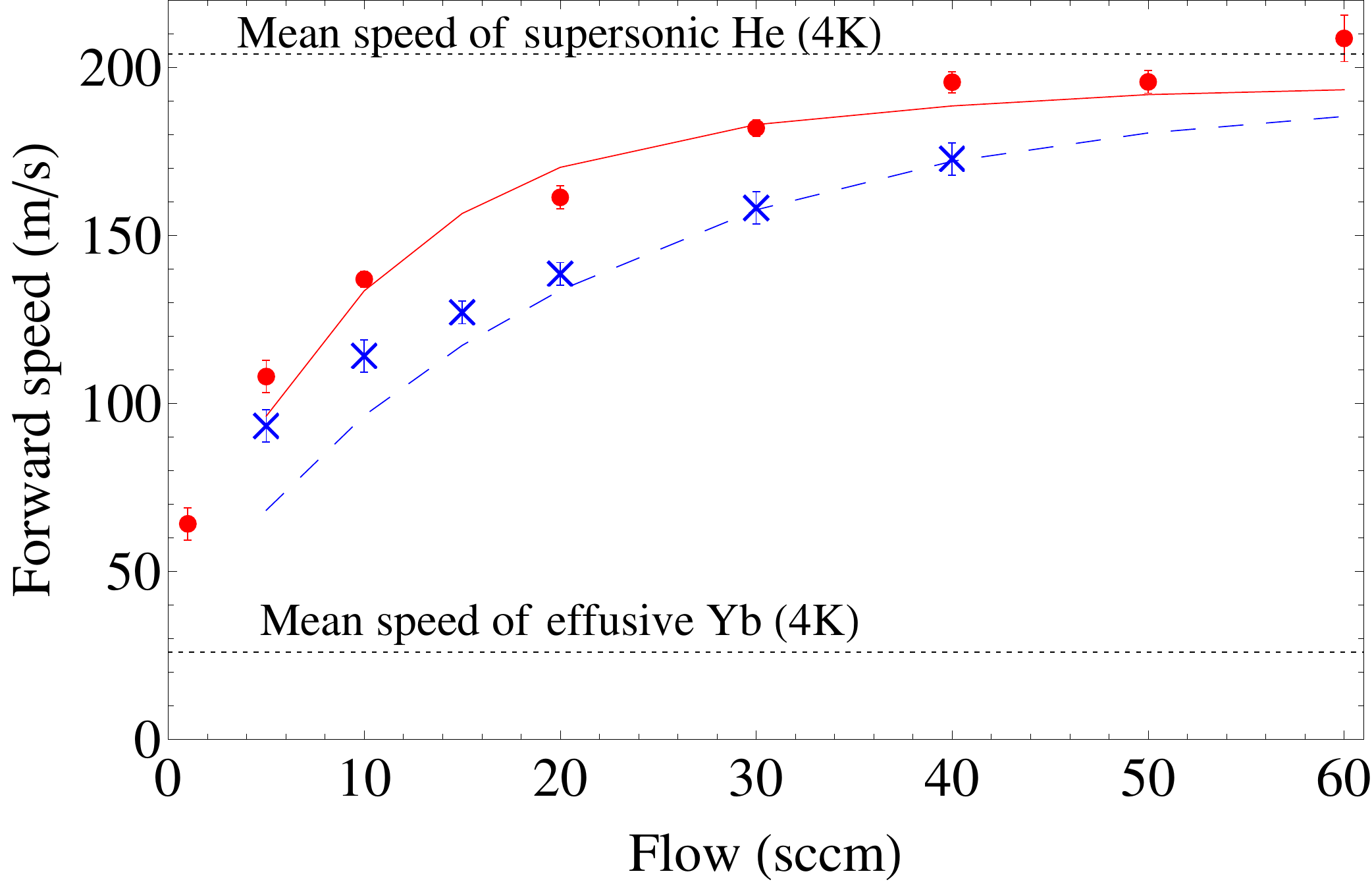}
	\end{center}
	\caption{Forward speed of the atomic beam versus flow, both measured (points) and simulated (lines). For the measurements, the ablation energy is 134\,mJ. Circles, solid line: 0.75\,mm x 4\,mm slit; Crosses, dashed line: 1\,mm x 8\,mm slit. The dotted lines show the mean speed of an effusive Yb beam and a supersonic He beam, both from a cell at 4\,K, and are the lower and upper limits to the speed.}
	\label{fig:vfgraph}
\end{figure}

We determine the mean forward speed of the atomic beam from the Doppler shift between spectra recorded with counter-propagating and perpendicular probe beams. This shift is calibrated by recording the transmission of a low-finesse optical cavity with a free spectral range of 150\,MHz. Figure \ref{fig:vfgraph} shows the forward speed measured at the lower detector versus flow for the two apertures. Here, the ablation power was 134\,mJ. The measured speed is averaged over the entire molecular pulse. For low helium flow we observe speeds as low as 65\,m/s, and as the helium flow is increased, the speed approaches the supersonic speed limit for helium at 4\,K, 204\,m/s. The velocity of an effusive beam of Yb at 4\,K is 26\,m/s, but we do not observe speeds this low. At the low flow rates required to reach the effusive regime, the density of helium in the cell is too low to stop the ballistic expansion of the Yb from the ablated target, and so almost all the atoms hit the walls.

Figure \ref{fig:vfgraph} also shows the prediction for the speed according to the flow dynamics simulation. Here, we assume that the Yb speed is simply that of the bulk helium flow out of the aperture. This makes sense because there are many collisions between the He and the Yb in the vicinity of the aperture, but hardly any collisions in the much lower density region downstream of the aperture. The Yb is entrained in the bulk flow of helium and so obtains the speed of this bulk flow, plus a small additional component due to its own thermal speed. With this assumption, we find that the predictions of the model agree well with the experimental results over the entire range of flow, particularly for the smaller slit. In both the experiment and the model, the smaller Yb slit gives a higher speed for the same flow rate. This is because, for the same flow rate, the flow velocity is higher when the aperture is smaller. For the larger slit, the model underestimates the speed at low flow, probably because the thermal component of the Yb becomes increasingly important when the flow velocity of the helium is low.

\begin{figure}[tb]
	\begin{center}
		\includegraphics[width=0.9\linewidth]{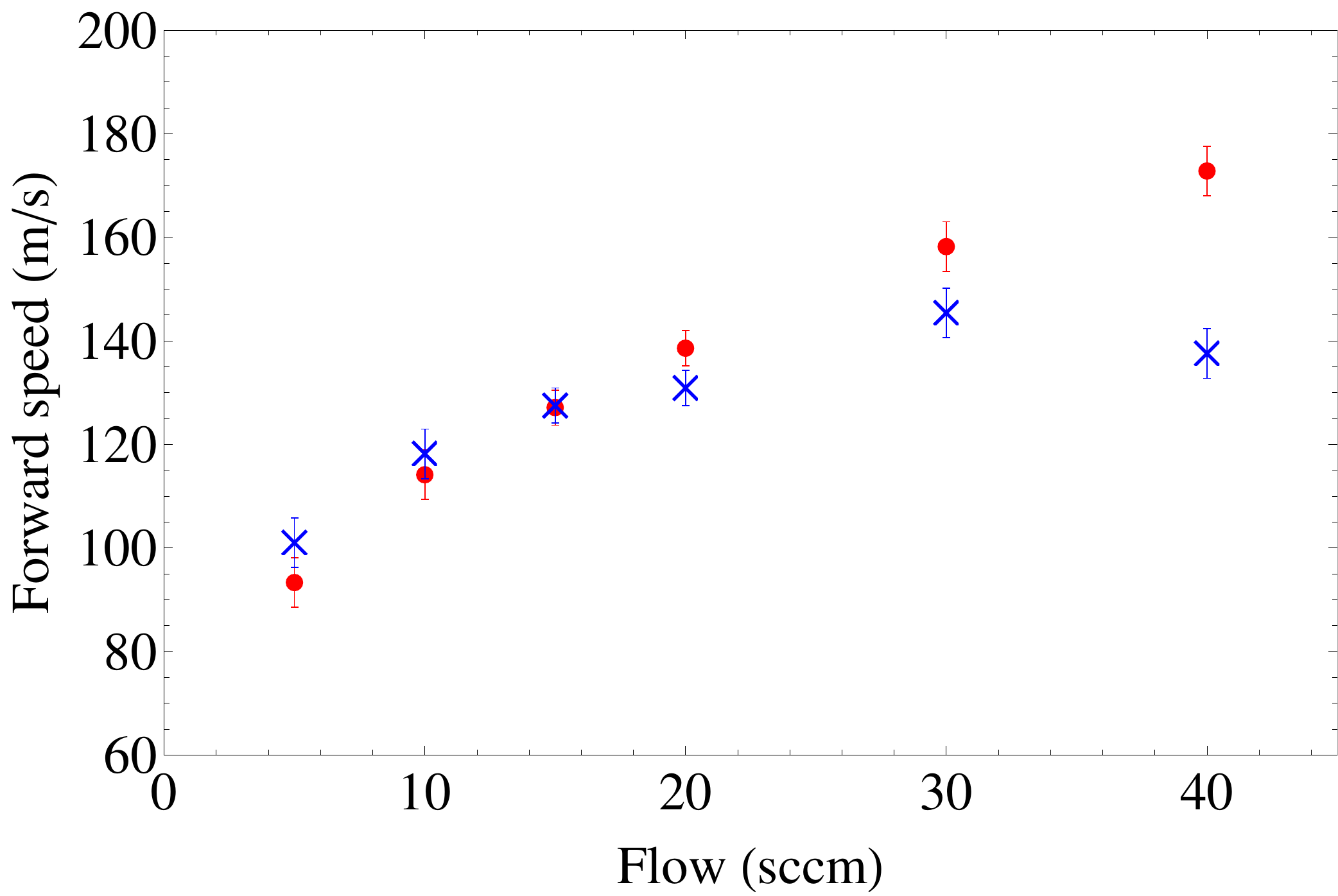}
	\end{center}
	\caption{Forward velocity as a function of flow of the beam extracted from the 1\,mm x 8\,mm slit, as measured at two detectors positioned 23\,mm (circles) and 132 mm (crosses) from the slit. The ablation energy is 134\,mJ.}
	\label{fig:velftwopmts}
\end{figure}

For both slit apertures we measure the speed of the beam at both the lower and upper detectors (see Fig.\ref{fig:Setup}). Figure \ref{fig:velftwopmts} shows the speed measured at the two detectors for the 1\,mm x 8\,mm slit. At low flow the speeds are the same, but at high flow the speed is smaller at the upper detector than it is at the lower detector. We suppose that at the highest flows, the background pressure of helium gas is high enough that the mean free path of the Yb atoms is shorter than the distance between the two detectors. Collisions with this background gas reduce the speed. The two detectors measure the same flux and temperature at low flow, but at high flow the flux at the upper detector is reduced (by 36\% at 40\,sccm) and the longitudinal temperature at the upper detector is increased (to 12\,K at 40\,sccm). These observations also support the idea that there are some collisions between the two detectors. We also observe the same trends for the 0.75\,mm x 4\,mm slit.

\begin{figure}[tb]
	\begin{center}
		\includegraphics[width=0.9\linewidth]{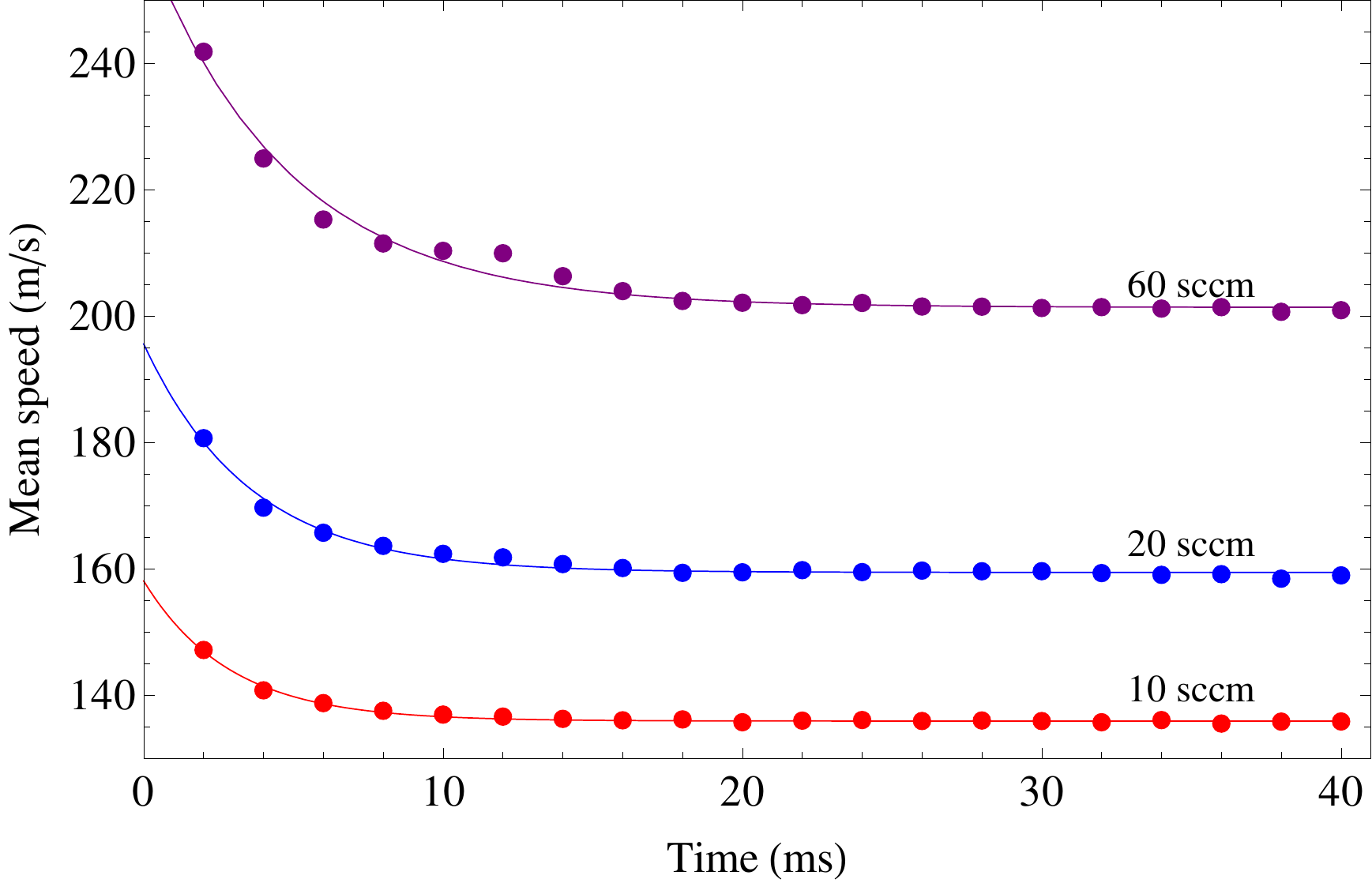}
	\end{center}
	\caption{Mean Yb speed as a function of arrival time at the lower detector, for the 0.75\,mm x 4\,mm slit and for flows of 10, 20 and 60\,sccm. The ablation energy is 134\,mJ. Lines are fits to an exponential decay model.}
	\label{fig:veltime}
\end{figure}

Figure \ref{fig:veltime} shows the measured forward velocity as a function of the arrival time at the lower detector, for some selected flows. The speeds are determined from the Doppler shift with the time-of-arrival profile integrated over time intervals of 0.05\,ms. We see that the velocity is higher at early arrival times. We fit the data to an exponential decay model, $v(t) = v_{f} + v_{e}\exp(-t/\tau)$, where $v_f$ is the final velocity, $v_f + v_e$ is the initial velocity, and $\tau$ is the time constant. We find that the excess velocity $v_e$ increases with flow from $26 \pm 6$\,m/s at 5\,sccm to $59 \pm 3$\,m/s at 60\,sccm. The time constants also increase with flow, from $2.2 \pm 0.4$\,ms at 5\,sccm to $4.8 \pm 0.3$\,ms at 60\,sccm. These results are consistent with the observation that the He and Yb are heated by the ablation pulse and then cool down as they re-thermalize with the cell walls, as discussed in section \ref{Sec:TemperatureInside}. For flows of 5 and 10\,sccm, the time constants for the velocity and the temperature are consistent with one another, but for higher flows the time constant for the velocity is smaller than for the temperature by a factor of 3--4. This may be because the speed is sensitive only to the conditions in the flow column whereas the temperature is averaged across a column perpendicular to the flow.

\subsection{Translational temperature of the beam.}

\begin{figure}[tb]
	\begin{center}
		\includegraphics[width=0.9\linewidth]{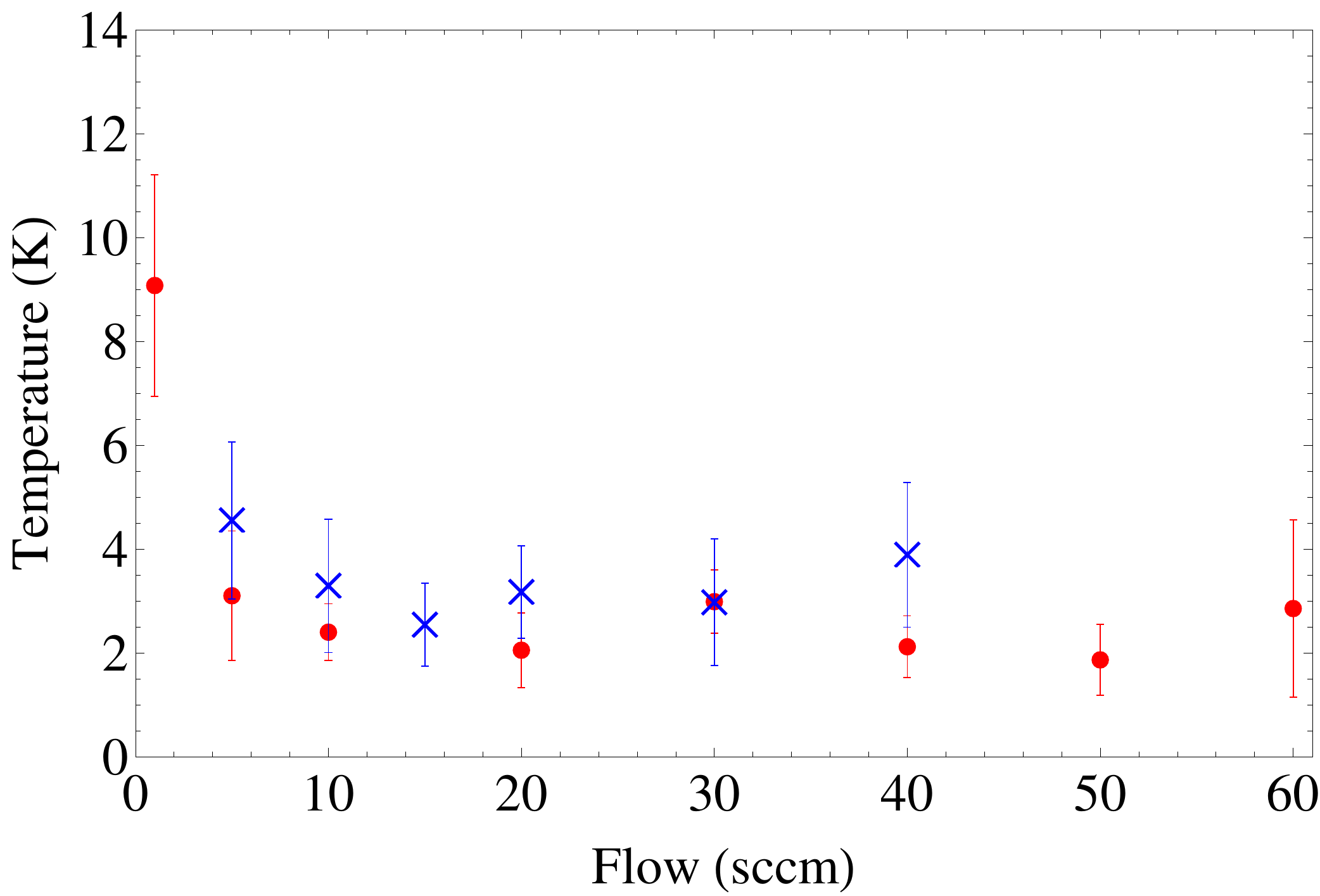}
	\end{center}
	\caption{Longitudinal translational temperature of the atomic beam for various helium flows. The ablation energy is 134\,mJ. Circles: 0.75\,mm x 4\,mm slit; Crosses: 1\,mm x 8\,mm slit. }
	\label{fig:tempL}
\end{figure}

The Doppler width of the fluorescence spectrum induced by the counter-propagating laser beam gives the longitudinal translational temperature of the atomic beam. Figure \ref{fig:tempL} shows this temperature as a function of flow for the two slits. For flows $\sc{F}\ge 5$\,sccm we find that the translational temperature is independent of the flow, with a weighted mean of $2.4 \pm 0.3$\,K for the 0.75\,mm x 4\,mm slit and $3.2 \pm 0.4$\,K for the 1\,mm x 8\,mm slit. The atoms in the beam are colder than the cell temperature showing that they are cooled as they expand through the slit. The expansion cooling is effective even with a 5\,sccm flow where the Reynolds number is only about 10. Cooling to temperatures below the cell temperature has also been observed by others \cite{Barry(1)11, Hutzler(1)11}

\subsection{Divergence of the beam.}

The transverse velocity spread was obtained from the Doppler width of the absorption profiles measured 23\,mm from the cell aperture, with the probe beam propagating parallel to the long dimension of the slit. These measurements, along with the measurements of the forward speed, are used to determine the divergence angle of the atomic beam. We define the full divergence angle as $\Delta\theta = 2 \tan^{-1}(\Delta v_{\text{FWHM}}/(2 v_f))$ where $\Delta v_{\text{FWHM}}$ is the full width at half maximum of the transverse velocity distribution, and $v_f$ is the mean forward speed. The divergence of the extracted beam is expected to be smaller in the intermediate flow regime than in either the effusive or fully hydrodynamic regimes. This is because, for intermediate flow, there are many collisions in the immediate vicinity of the aperture where the flow is highly collimated, whereas collisions downstream of the aperture, which would tend to broaden the transverse velocity distribution, are rare. In the limiting case, we suppose that the forward speed is boosted to the supersonic speed of the helium, whereas the transverse speed distribution retains the thermal distribution of the species at the cell temperature. In this limiting case the divergence would be
\begin{equation}
\Delta\theta_{\text{min}} = \sqrt{\frac{8 \ln 2}{5}\frac{M_{\text{b}}}{M_{\text{s}}}}
\end{equation}
where $M_{\text{b}}$ is the mass of the buffer gas atom and $M_{\text{s}}$ is the mass of the heavier species atom. In our case of Yb carried in He, the minimum divergence is $\Delta\theta_{\text{min}} = 9$\,degrees.

\begin{figure}[tb]
	\begin{center}
		\includegraphics[width=0.9\linewidth]{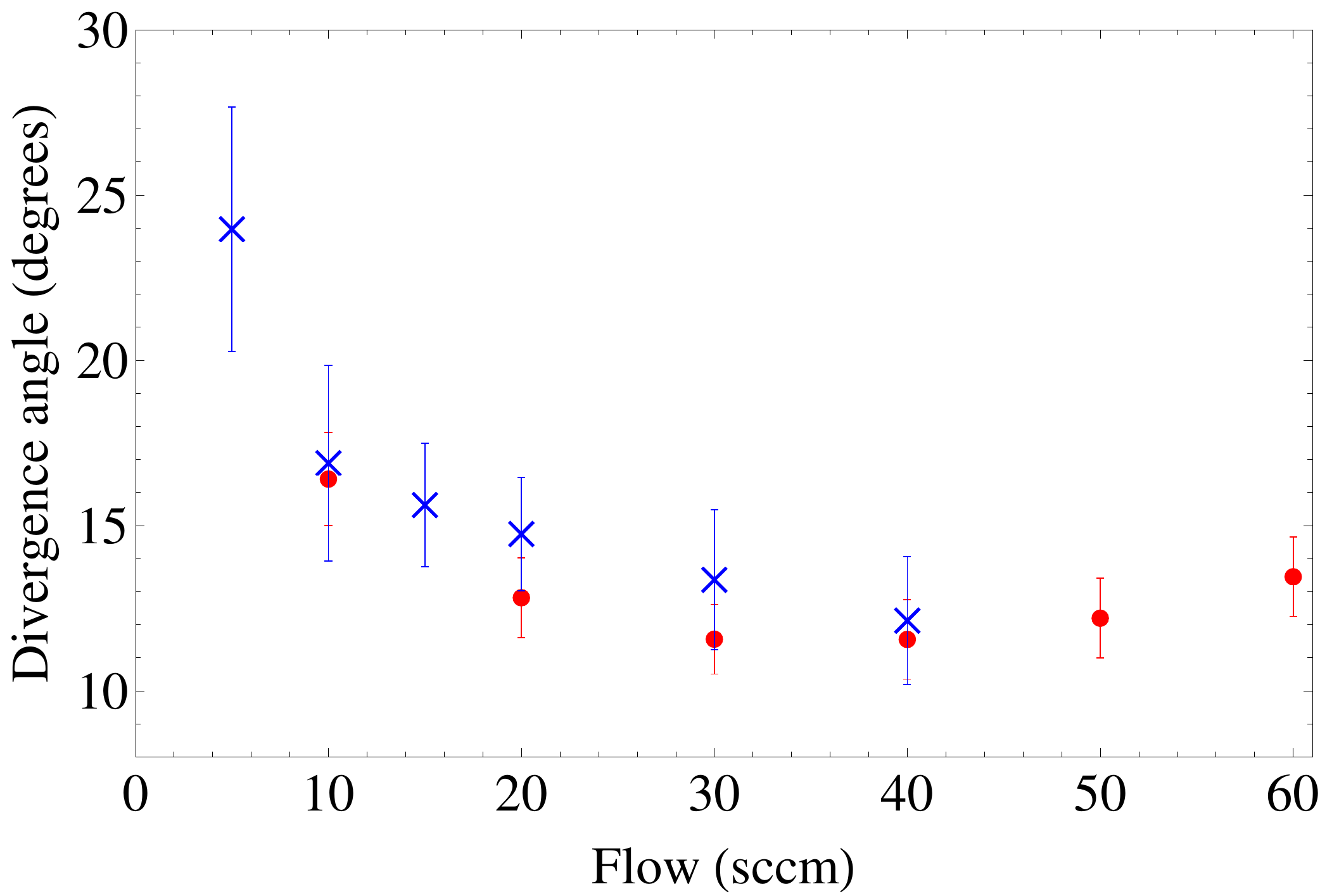}
	\end{center}
	\caption{Measured divergence angle, $\Delta \theta$, as a function of flow, for the 0.75\,mm x 4\,mm slit (circles) and the 1\,mm x 8\,mm slit (crosses). The ablation energy is 134\,mJ.}
	\label{fig:divgraph}
\end{figure}

Figure \ref{fig:divgraph} shows the measured divergence as a function of the flow for the two apertures. As expected from the discussion above, the divergence decreases as the gas flow increases. At a flow of 30-40\,sccm, the divergence angle reaches its minimum value of about 12 degrees, only a little higher than the value of $\Delta\theta_{\text{min}}$ predicted above. The reduction in divergence correlates with the increase in forward speed up to this flow. For higher flows there is only a small further increase in the forward speed and the divergence should increase again as downstream collisions start to broaden the transverse speed distribution. We did not take data at high enough flows to observe this clearly. The highly collimated beam produced by operating in this intermediate flow regime is favourable for most experiments where a high flux is required far downstream of the cell. The minimum beam divergence measured here is 2-3 times smaller than that measured for molecular beams of SrF \cite{Barry(1)11} and of ThO \cite{Hutzler(1)11}, for similar values of the Reynolds numbers. This is probably because of differences in the aspect ratio of the apertures. For the SrF and ThO divergence measurements, the transverse size of the aperture was much larger than the aperture thickness. In our measurements, the thickness of the aperture is about the same as its width. This tends to channel the helium gas exiting the cell, producing a more collimated helium beam, which is then reflected in the low divergence we measure for the Yb beam.

\subsection{Extraction efficiency.}

\begin{figure}[tb]
	\begin{center}
		\includegraphics[width=0.9\linewidth]{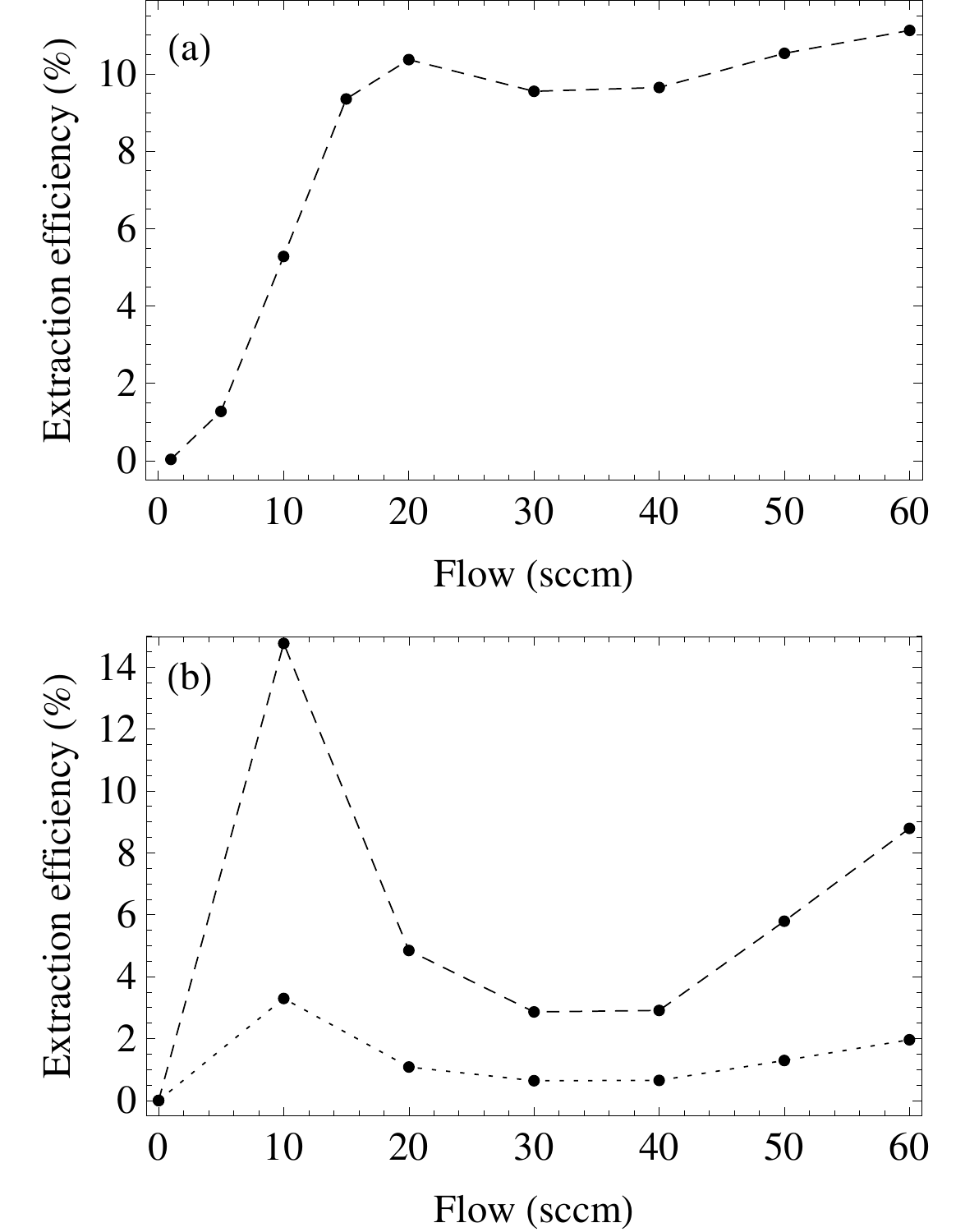}
	\end{center}
	\caption{(a) Simulated extraction efficiency for the 0.75\,mm x 4\,mm slit. (b) Upper and lower limits to the measured extraction efficiency for the 0.75\,mm x 4\,mm slit and for an ablation energy of 79\,mJ. For the lower limit, we assume a uniform distribution throughout the cell. For the upper limit we assume the atoms are confined to the region of the probe laser.}
	\label{fig:ExtractionEfficiency}
\end{figure}

The extraction efficiency is the total number of atoms emitted into the beam divided by the total number of atoms inside the cell. Figure \ref{fig:ExtractionEfficiency}(a) shows the simulated extraction efficiency, obtained by normalizing the total number of Yb atoms that leave the cell during an 80\,ms period to the total number present in the initial distribution. In this simulation, the initial distribution was a sphere of radius 5\,mm offset from the centre of the cell by 8\,mm in the direction of the target. Up to a flow rate of 20\,sccm the extraction efficiency increases with increasing flow as more atoms are swept up in the flow instead of diffusing to the cell walls. There is then a small dip in the extraction efficiency at around 30--40\,sccm, which is due to a narrowing of the flow column and the formation of vortices which trap molecules inside the cell. For higher flows the narrowing of the flow column ceases and we see the extraction efficiency slowly increasing again. For this initial distribution, the simulation predicts an extraction efficiency of about 10\% for flows between 15 and 50\,sccm. The simulated extraction efficiency is sensitive to the initial distribution of atoms in the cell, but the general trend is the same: a rapid increase in efficiency as the flow is increased up to 20\,sccm, then a much more gradual increase for higher flows. When the same spherical distribution is located in the middle of the cell, the efficiency at 60\,sccm increases to 45\%. Conversely, if the initial distribution is confined to a corner of the cell, far from the flow column, the efficiency at 60\,sccm is only 0.7\%.

To measure the extraction efficiency we determine the number of molecules in the beam and in the cell. We use the laser-induced fluorescence signal at the lower detector to estimate the number of atoms per unit solid angle in the beam. The fluorescence detection efficiency is calibrated by comparing absorption and fluorescence measured simultaneously at the lower detector. Using the divergence angles shown in Fig.\,\ref{fig:divgraph}, we obtain the total number of atoms in the beam. We use the absorption measured through the cell to estimate the number of atoms in the cell. The absorption changes with time, and we take the maximum value. In relating the measured absorption to the atom number density, we account for Doppler broadening, pressure broadening, the saturation of the atomic transition, and the change in laser intensity as the probe beam propagates through the sample. We measure the mean atom density in the region occupied by the probe laser, which has a diameter of 16\,mm. From this mean density we estimate a lower limit to the extraction efficiency by assuming that the density is the same throughout the volume of the cell, and we estimate an upper limit by assuming that all of the atoms are located within the volume of the probe laser. Figure \ref{fig:ExtractionEfficiency}(b) shows these upper and lower limits as a function of flow for the 0.75\,mm x 4\,mm slit. For low flows we expect the cell to be filled approximately uniformly and the extraction efficiency to follow the lower curve. For higher flows the high density of buffer gas confines the atoms, as we see in our absorption images, and we expect the extraction efficiency to follow the upper curve more closely. The overall trends predicted by the simulation are observed in the experiment. In particular, the dip around 30--40\,sccm is observed. The absolute values are smaller than predicted, particularly in the region of the dip, which perhaps indicates that the circulation regions are even less conducive to good extraction than the simulation suggests.

Previous work has shown that extraction efficiency depends strongly on the ratio $\gamma = \tau_{D}/\tau_{p}$ \cite{Patterson(1)07}. The extraction efficiency can reach high values when $\gamma > 1$. For our cell geometry, the extraction efficiency varies strongly through the cell volume, being high in the flow column and low everywhere else. Nevertheless, it is useful to consider the average value of $\gamma$ obtained using equations (\ref{equ:pump}), (\ref{equ:diff}) and (\ref{equ:diffcoefficient}). These give $\gamma \simeq 0.01 F$ for our cell, where $F$ is the flow in sccm. At our highest flow of 60\,sccm, $\gamma$ reaches 0.6 and our measured extraction efficiency lies between 2 and 8\%. For this value of $\gamma$, the extraction efficiency measured in reference \cite{Patterson(1)07}, where the cell geometry is similar to ours, was about 10\%. This was found to increase to about 40\% for $\gamma > 1$. For SrF, extraction efficiencies close to 50\% are reported for values of $\gamma$ between 0.01 and 0.5 \cite{Barry(1)11}. Here however, the extraction efficiency was determined by comparing the density 1\,mm upstream and 1\,mm downstream of the exit aperture. The extraction efficiency measured this way is high since most molecules this close to the exit aperture would be expected to leave the cell. If the total number of molecules in the cell were used instead of the number at the aperture, we would expect the extraction efficiency to be lower. In reference \cite{Hutzler(1)11} the extraction efficiency for ThO molecules is about 20\% once the flow is high enough to ensure that $\gamma > 1$. In this work, the cell had a diameter of only 13\,mm, and the exit aperture was a square with side length between 1.5 and 4.5\,mm. This geometry gives a more uniform column of flow throughout the cell and so gives higher overall extraction efficiency. In our geometry, the thickness of the exit slit may also influence the extraction efficiency. The slits are as thick as they are wide. This tends to channel the flow, producing a well-collimated beam, but it also chokes the flow and encourages the re-circulation regions to form, which limits the extraction efficiency.

\section{Summary and conclusions}

Our combined experimental and theoretical study adds to the growing understanding of beam formation from a cryogenic buffer gas cell. The geometry of the cell determines how the gas flows from inlet to outlet, and this has a large effect on the extraction of atoms and molecules loaded into the cell. Those loaded into the region of high flow are swept out rapidly and efficiently, forming a peak in the arrival-time distribution with a typical width of 5--10\,ms. Atoms that are loaded into the stagnant regions or regions where the gas is circulating are extracted slowly and inefficiently, and produce a long tail in the arrival time distribution, up to 100\,ms in duration. To extract the maximum number of atoms or molecules, there should be a sufficiently high flow through the whole volume of the cell. The transit time of the helium through the cell should be much shorter than the diffusion time to the walls in all regions of the cell. The density of the helium and the size of the cell need to be sufficient to stop the ballistic expansion of the ablation plume before it reaches the cell walls. We measure extraction efficiencies of a few percent.

Atoms and molecules produced by laser ablation thermalize rapidly with the helium gas in the cell, but the ablation process raises the temperature of the helium, typically by about 10\,K. The helium then re-thermalizes to the temperature of the cell walls with the timescale set by diffusion of heat to the walls, between 5 and 20\,ms for the helium densities used in these experiments. We observe pressure broadening of the Yb 556\,nm transition, and have determined the pressure broadening cross-section at 5\,K to be $(1170 \pm 50)\times 10^{-20}$\,m$^{2}$. The speed of the atomic beam increases with increasing gas flow. For a 0.75\,mm x 4\,mm exit aperture, the speed is 65\,m/s for a flow of 1\,sccm, and increases with flow until it reaches 204\,m/s, which is the supersonic speed of He at 4\,K. Our measurements agree well with the numerical model where we assume that the Yb atoms acquire the bulk speed of the helium leaving the aperture. The atoms that leave the cell first are faster than those that leave later, and this we attribute to the heating of the helium by the ablation process. The atoms are cooled below the cell temperature as they expand out of the aperture. For all flows greater then 5\,sccm, the translation temperature of the beam is 2-3\,K. The beam is exceptionally well collimated, having a full divergence angle of just 12 degrees for flows of 30-40\,sccm. The low divergence is as expected from a simple model where all collisions occur in the immediate vicinity of the aperture where the gas flow is highly collimated.
\vspace{0.5cm}

\noindent\textbf{Acknowledgements}

We are grateful to Jon Dyne, Steve Maine and Valerijus Gerulis for their expert technical assistance, and to Danny Segal for fruitful discussions. This work was supported by the Royal Society, and by the EPSRC.

\end{document}